# Multi-cavity ultrastable laser towards 10⁻¹⁸

ZHAOYANG TAI,[1,2] LULU YAN,[1,2,†] YANYAN ZHANG,[1,2] PAN ZHANG,[1] XIAOFEI ZHANG,[1,2] WENGE GUO,[1] SHOUGANG ZHANG,[1,2] AND HAIFENG JIANG,[1,2,*]

[1]*Key Laboratory of Time and Frequency Primary Standards, National Time Service Center, Chinese Academy of Sciences, Xi'an 710600, China*
[2]*School of Astronomy and Space Science, University of Chinese Academy of Sciences, Beijing 100049, China*
†*joint first author*
*Corresponding author: haifeng.jiang@ntsc.ac.cn*

**In this letter, we demonstrate a technique of making an ultrastable laser referenced to a multi-cavity, corresponding to a lower thermal noise limit due to the larger equivalent beam size. The multi-cavity consists of several pairs of mirrors and a common spacer. We can stabilize the laser frequencies on these cavities, and average the laser frequencies with synthesizing technique. In comparison with a single cavity system, relative frequency instability of the synthesized laser can be improved by a factor of the squre root of the cavity number ($n$). In addition, we perform an experiment to simulate a two-cavity system. Experimental results show that frequency instability of the synthesized laser is improved by a factor of $\sqrt{2}$, and discrimination of the laser frequency instability, introduced by the process of lasers synthesizing, is negligible, and can reach a floor at low level 10⁻¹⁸ limited by noise of currently used signal generators. This technique is comparable with other techniques; thus, it can gain a factor of $\sqrt{n}$ on the frequency instability of an ultrastable laser to an unprecedented level.**

Ultrastable laser is vital component in many scientific applications including optical atomic clocks [1], gravitational wave detection with laser interferometry [2], dark matter detection[3], and verification of physics laws [4]. State-of-the-art ultrastable lasers, limited by the thermal noise effect, exhibit frequency instability at 10⁻¹⁷ level [5–7]. However, the quantum projection noise level of the best optical clocks has reached 10⁻¹⁸ @ 1 s level [8, 9]. With the best quantum noise limited optical atomic clocks, the newly proposed gravitational wave detection proposal becomes more realistic[10]. More stable laser systems are required.

The main contribution of the thermal effect comes from the mirror displacement ($\Delta L$) due to Brownian motion. According to the fluctuation dissipation theorem (FDT) [11], a mirror's thermal noise effect described as the double-sideband power spectrum is given in Ref. 12.

$$G_{\mathrm{mirror}}(f) = \frac{4k_B T}{\omega} \frac{1-\sigma^2}{\sqrt{\pi} E w_0}(\phi_{\mathrm{sub}} + \frac{2}{\sqrt{\pi}} \frac{1-2\sigma}{1-\sigma} \frac{d}{w_0}\phi_{\mathrm{coat}}) \quad (1)$$

where $k_B$ is Boltzmann's constant, $T$ the temperature, and $\omega=2\pi f$ the angular frequency; $\sigma$, $E$, and $\phi_{\mathrm{sub}}$ are the Poisson's ratio, Young's modulus and mechanical loss of the mirror substrate, respectively; $d$ and $\phi_{\mathrm{coat}}$ are the thickness and mechanical loss of the coating; $w_0$ is the beam radius of a Gaussian beam. The frequency instability of a cavity stabilized laser is determined by the length instability of the reference cavity; the mirror's thermal noise effect on laser can be written as:

$$\frac{\sqrt{S_\nu(f)}}{\nu} = \frac{\sqrt{2G_{\mathrm{mirror}}(f)}}{L} = \sqrt{\frac{4k_B T}{\pi^{3/2} f} \frac{1-\sigma^2}{E w_0 L^2}(\phi_{\mathrm{sub}} + \frac{2}{\sqrt{\pi}} \frac{1-2\sigma}{1-\sigma} \frac{d}{w_0}\phi_{\mathrm{coat}})} \quad (2)$$

where $S_\nu(f)$ is the frequency noise, $\nu$ the laser frequency, and the 2 in front of $G_{\mathrm{mirror}}(f)$ represents two mirrors of a cavity. This flicker noise corresponds to the relative frequency instability independent of averaging time, indicated by Allan deviation[13, 14]:

$$\sigma_A = \sqrt{2\ln 2 f} \frac{\sqrt{S_\nu(f)}}{\nu} = \sqrt{\ln 2 \frac{8k_B T}{\pi^{3/2}} \frac{1-\sigma^2}{E w_0 L^2}(\phi_{\mathrm{sub}} + \frac{2}{\sqrt{\pi}} \frac{1-2\sigma}{1-\sigma} \frac{d}{w_0}\phi_{\mathrm{coat}})} \quad (3)$$

Equation (3) contains all factors that determinate the thermal noise effect. Over the last decade, efforts have been made to reduce this effect by using long cavity [7], applying low mechanical loss materials [15–18], cooling the cavity [6, 17], or enlarging the beam size [9, 19]. It is difficult to have a cavity longer than half meter, because of the lack of such a bulky ultralow expansion (ULE) glass and the complexity for reducing the vibration sensitivity and temperature fluctuation. Up to now, the longest ultrastable cavity has been 48 cm [7]. Regarding the choice of the mirror substrate, ULE glass, fused silica, single-crystal silicon and special coatings have been investigated. Fused silica exhibits a smaller mechanical loss than ULE glass [15, 16]; single-crystal silicon has a larger Young's modulus than fused silica and ULE, and cavities entirely made of it have shown great thermodynamic performance [17]; and monocrystalline multilayer coatings have smaller mechanical loss than dielectric multilayer coatings [18]. Providing a cryogenic environment for the cavity is a straightforward way of reducing the thermal noise. Recently, a 21-cm-long silicon cavity at 124 K and a 6-cm long silicon cavity at 4 K exhibit a relative instability of $4\times10^{-17}$ and $1\times10^{-16}$ [5, 6]. Enlarging the laser beam size must be achieved by using a concave mirror with larger radius of curvature [9, 19]. However, a larger radius cavity leads to difficulties of mode matching due to high-order modes being excited more easily[20] .

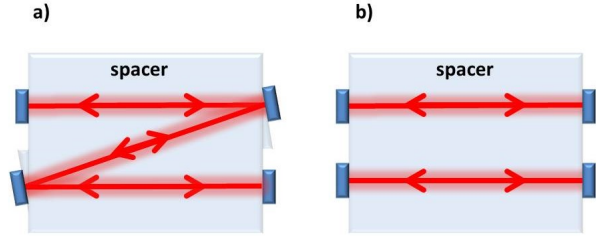

Fig.1.Diagrammatic sketch of (a) fold cavity and (b) muti-cavity.

It seems that scientists have considered all possible factors shown in Eq. (3), including cavity length, temperature, cavity materials, and beam size, to reduce the thermal noise effect of a two-mirror cavity. An alternative way is use of a fold-cavity, because its equivalent cavity length is longer. For example, a four-mirror fold-cavity [see Fig. 1(a)] can reduce the thermal effect by a factor of approximately $\sqrt{5/3}\approx0.75$, assuming that all mirrors have the same thermal noise level and that the tilt angle of light is small. However, this idea may not possible to realize due to difficulties in making and mounting the tilt mirrors. Considering that the photons bound back and forth between both sides of the spacer, we also consider that the fold cavity has a larger equivalent beam size. The question arises of why we do not mount several [$n$=2 in this case; see Fig. 1(b)] separated cavities onto a common spacer. If one can average the lengths of these cavities, the equivalent beam size is then enlarged $n$ times; consequently, the relative length instability gains $\sqrt{n}$ ($1/\sqrt{2}\approx0.71$ in this case), because the thermal noise on different cavities is not coherent. One can expect that a 21-cm-long silicon cavity operating at 4 K rather than at 124 K should exhibit a relative length instability 5–6 times lower than $4\times10^{-17}$; considering that some technical noise may appear at this level, the laser instability should be close to $1\times10^{-17}$. If one can make a similar multi-cavity system to further improve the laser instability, $10^{-18}$ is definitely possible.

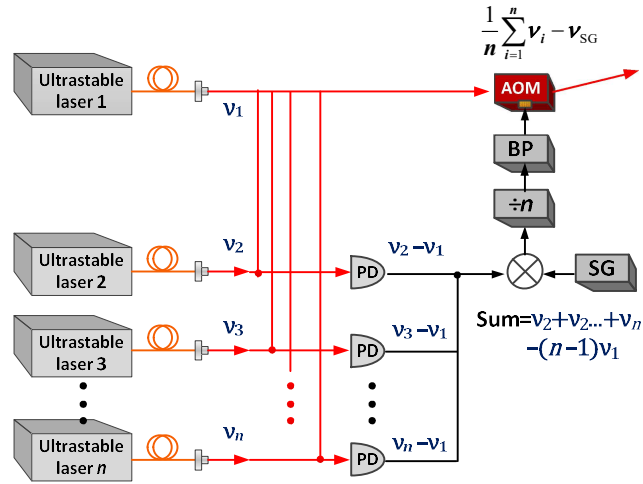

Fig. 2. Schematic of synthesizing lasers into one by averaging their frequencies, PD, photodiode; SG, signal generator; ÷$n$, frequency divider; BP, band-pass filter; AOM, acoustic-optical modulator.

Now the question becomes how to average the length of cavities, i. e., frequency of lasers, performing a laser with frequency of

$$\nu_{1,n} = \frac{1}{n}\sum_{i=1}^{n}\nu_i \qquad (4)$$

where $n$ is the number of cavities, $\nu_i$ the laser frequency. Direct optical synthesizing by the nonlinear effect is difficult, especially for the dividing process. Fortunately, we can subtract between lasers by using a photodetector, and sum the laser and radio frequency (RF) by using an acoustic-optical modulator (AOM). Thus, we can perform such an optical frequency synthesizing by employing mature RF techniques. Figure 2 shows the schematic of laser synthesizing. Lasers should be frequency-stabilized onto cavities first; then, we can combine one laser (laser 1) with all other lasers, and direct them to photodetectors in which heterodyne signals ($\nu_1$–$\nu_2$, $\nu_1$–$\nu_3$, …) can be produced. Frequency summation by RF mixers yields the
the sum frequency of $\nu_2$+…+$\nu_n$–($n$-1)$\nu_1$ . Then, one can frequency-divide this signal to obtain [$\nu_2$+…+$\nu_n$–($n$-1)$\nu_1$]/$n$. To match the resonant frequency of the AOM, an additional frequency shift ($\nu_{SG}$) can be provided by a low-noise signal generator. Finally, the synthesized laser $\frac{1}{n}\sum_{i=1}^{n} \nu_i - \nu_{SG}$ is produced at the output of the AOM.

To verify this technique, we perform an experiment to simulate a two-cavity system by using two identical but independent cavity-stabilized lasers. Figure 3 shows the experimental setup. The ultrastable lasers, based on commercial 1555-nm lasers, exhibit a frequency instability of $7\times10^{-16}$ and a linewidth of ~100 mHz, mainly limited by the thermal noise effect of 10-cm ULE cavity [21]. The beatnote between two lasers ($\nu_1-\nu_2$=1.70 GHz) is detected by a photodiode (ET3000A, Electro-Optics Technology) and sent into a frequency divider with a factor of 2 (based on MC10EP139, ON Semiconductor), and then we obtain the divided frequency ($\nu_1-\nu_2$)/2 at 0.85 GHz. Next, we down-convert this signal by mixing with a reference frequency $\nu_{SG1}$ (0.74 GHz) provided by a signal generator (SG382, Stanford Research Systems) to 110 MHz, which is the resonant frequency of the tank circuits of the AOM (MGAS110-A1, AA Opto Electronic). At the same time, the other part of laser from $U_1$ passes though the AOM via Bragg diffraction [22]. By picking up the negative first-order diffraction light, we can obtain the synthesized laser with a frequency of

$$\nu_{1\&2} = \nu_2 + \nu_{AOM} = (\nu_1 + \nu_2)/2 - \nu_{SG}. \quad (5)$$

We have checked that the beat frequency $\nu_1-\nu_2$ is positive, however, this is not always the case. Actually, if the beatnote frequency ($\nu_1-\nu_2$) is negative, we can produce the target frequency item of ($\nu_1+\nu_2$)/2 from the positive first-order diffraction light. Note that we use the beatnote of $\nu_1-\nu_2$ at 1.70 GHz instead of a low frequency (e.g., 200 MHz) to avoid effects by stray noises due to laser frequency stabilization and fiber noise cancellations.

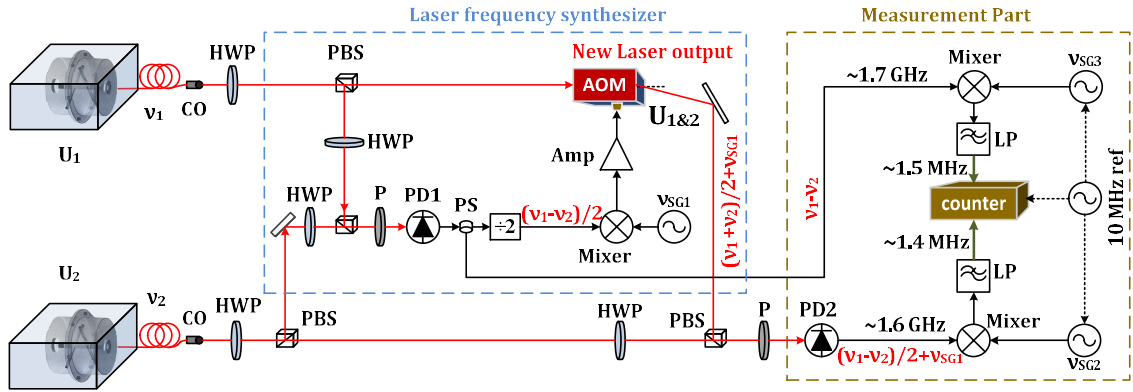

Fig. 3 Experimental setup of an ultrastable laser with two cavities. U, ultrastable laser; CO, collimator; PBS, polarization beam splitter; HWP, half-wave plate; P, polarizer; PD, photodiode; ÷2, frequency divider; SG, signal generator; LP, low pass filter; PS, RF signal power splitter. red line, optical signal; black line, electronic signal.

To measure the additional noise introduced by the optical frequency synthesizing process, we produce another beatnote between $\nu_1$ and $\nu_{1\&2}$, i. e., ($\nu_1-\nu_2$)/2−0.85 GHz, as shown in Fig. 3. We down convert two optical beatnotes to the operational range of a multichannel frequency counter (FXQE80, K+K Messtechnik) by mixing them with reference frequencies given by signal generators. Since the optical beatnotes are significantly noisier than the RF signals, we expect that the instability of these two beatnotes has a relation of factor 2. The discrimination of such a ratio between two beatnotes is the noise effect due to the frequency synthesizing. Note that we have synchronized SG2, SG3 and the frequency counter to a common 10MHz reference frequency.

As shown in Fig. 4, the relative frequency instability of two independent lasers is at $1\times10^{-15}$ level for 1–20 s integration time, while that of one laser against the synthesized laser is at $5\times10^{-16}$ level, as expected; actually. This also indicates frequency instability of the synthesized laser, because $\nu_1$ and $\nu_2$ are independent, so that the deviations of ($\nu_1-\nu_2$)/2 and ($\nu_1+\nu_2$)/2 are same. The discrimination from ratio 2 is $7\times10^{-17}$ @ 1 s and rolls down to $1\times10^{-17}$ @16 s. It is clear that the additional noise indicated by the discrimination is negligible. This high discrimination level is due to the fact that the laser beatnote is noisy, and that the frequency counter is not synchronized perfectly between these two channels. In order to verify this, we replace the beatnote signal from PD1 with a low noise 1.7 GHz RF signal. Then we obtain three noise floor levels corresponding to the three upper curves. All the noise floors are in the range of $10^{-18}$ and $10^{-19}$. Since frequency instability no longer agrees with the ratio 2, we conclude that the noise is mainly attributed to the noise introduced by signal generators. Whatever the cause, the additional noise effect of synthesizing (the discrimination floor) is already as low as $10^{-18}$. A lesser noise effect can be realized with better RF signals to achieve frequency synthesizing. In a crucial case, this effect could be minimized by using photonic microwave generation technique, yielding ultra-low-noise microwave signals in the $10^{-16}$ level, which is a few orders of magnitude better than the RF signals in this case [23].

As mentioned above, a space-saving solution is to place high finesse cavities in a common spacer, e. g., a cubic cavity body with cavities orthogonal to each other in the mid horizontal plane. Such a force-insensitive cubic cavity body with only one cavity was designed several years ago [24]. With a long cavity, it is also possible to have a stable multi-cavity with low vibration sensitivity by designing the cavity shape and support mechanism. In fact, it is not necessary to optimize each cavity, but rather to optimize the sum of these cavity lengths. It is worth noting that other techniques, such as the use of spectral-hole burning [25, 26], can benefit from this approach to improve a laser's frequency instability by averaging frequencies of all lasers.

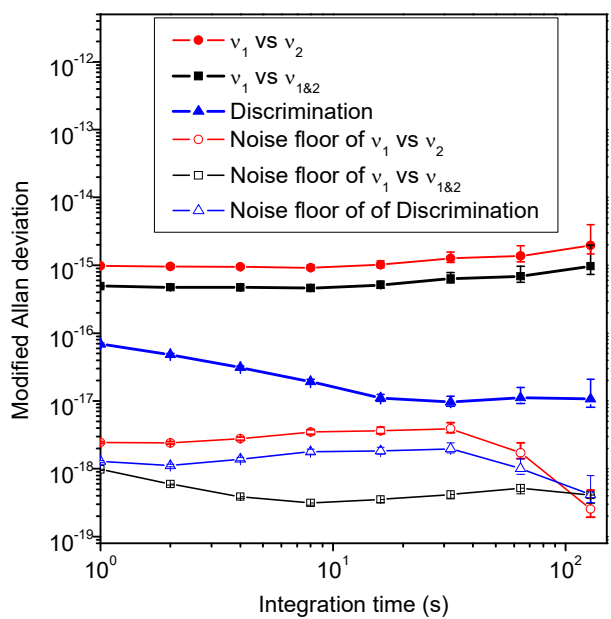

Fig. 4. Relative frequency instability of lasers and their noise floors.

In summary, we have demonstrated a proposed approach for improving the frequency instability of ultrastable lasers by using a multi-cavity design. In addition, we have performed an experiment to simulate a two-cavity system. Our result shows that frequency instability of the synthesized laser is improved by a factor of $\sqrt{2}$, and the additional frequency instability induced by frequency synthesizing is at low level of $10^{-18}$. This noise effect is attributed to the noise of RF signal generators. With better RF signal generators or by using a photonic microwave generation technique, this effect can be further reduced. Based on a multi-cavity (2–4 cavities) design, one can further improve frequency instability of current ultrastable lasers by a factor of 1.4–2, up to $10^{-18}$ or even better.